\documentclass[a4paper]{article}
 \usepackage{graphicx}
 \usepackage{amsmath}
 \usepackage{hyperref}
 \usepackage{color}

\begin{document}
 
 \begin{center}
 
 {\bf \Large Clustering in random line graphs}\\[5mm]

 {\large Anna Ma\'nka-Kraso\'n, Advera Mwijage$^*$ and Krzysztof
 Ku{\l}akowski}\\[3mm]

 {\em
 
 Faculty of Physics and Applied Computer Science, AGH University of
 Science and Technology, al. Mickiewicza 30, PL-30059 Krak\'ow,
 Poland\\
 
 $^*$On leave from Mbeya Institute of Science and Technology, P.O.Box
 131 Mbeya, Tanzania
 
 }

 
 {\tt kulakowski@novell.ftj.agh.edu.pl}

 \bigskip
 
 \today
 
 \end{center}
 
 \begin{abstract}

We investigate the degree distribution $P(k)$ and the clustering coefficient $C$ of the line graphs constructed 
on the Erd\"os-R\'enyi networks, the exponential and the scale-free growing networks. We show that the
character of the degree distribution in these graphs remains Poissonian, exponential and power law, respectively,
i.e. the same as in the original networks. When the mean degree $<k>$ increases, the obtained clustering coefficient
$C$ tends to 0.50 for the transformed Erd\"os-R\'enyi networks, to 0.53 for the transformed exponential networks and 
to 0.61 for the transformed scale-free networks. These results are close to theoretical values, obtained with the model 
assumption that the degree-degree correlations in the initial networks are negligible.

 \end{abstract}
 
 \noindent
 
 {\em PACS numbers:} 64.60.aq; 02.10.Ox; 05.10.Ln
 
 \noindent
 
 {\em Keywords:} line graphs; random networks; clustering coefficient; degree distribution

 \bigskip
 \section{Introduction}
 
 The science of networks is indeed a new kind of science \cite{b0} for
 its interdisciplinary character and its explosive development;
 for some recent monographs we refer to \cite{b1,b2,b3,b4,b5,b6,b7,b8}.
 The list of applications of networks contains examples from physics,
 informatics, biology and social sciences. Basic characteristics of
 networks are the degree distribution $P(k)$ and the clustering
 coefficient $C$. The degree of a given node is the number of other
 nodes connected to that node; the clustering coefficient measures
 the probability that two neighbours of a given node are connected to
 each other. As it was indicated only recently by Mark Newman
 \cite{repr},
 many real networks show a high clustering coefficient, usually some
 tens of percent. On the contrary to this fact, model random networks
 show rather low $C$, unless a special procedure is applied to enhance
 it. Examples of such procedures are described in
 \cite{e1,e2,e3,e4,e5};
 we owe this list again to \cite{repr}. The idea is to enhance $C$
 gradually by linking nodes which are neighbours of the same node. When
 a node has three neighbours, it is convenient to replace it by a
 triangle \cite{my}; the trick is similar to the star-triangle or
 star-delta transformation. The latter has been used also to construct 
 Apollonian networks with a high values of the clustering coefficient \cite{nb},
and to prove some theorems in the theory of percolation \cite{per}.\\

The transformation from a graph $G$ to its line graph $L(G)$ \cite{wiki} used here can 
be seen as a simple reformulation of the star-triangle transformation. 
The latter converts a subgraph $Y$, i.e. a node with three neighbours, into a
 subgraph $\Delta$, where these three neighbours are linked
 to each other and the central node is deleted. In the transformation $G \to L(G)$ the
 same original subgraph $Y$ is converted also into a graph
 of three nodes linked to each other.  The idea of the line graph, known also as
edge graph \cite{zbrd}, is to
 convert all links into nodes \cite{wiki}. In this way a new network appears,
 where the number of nodes is equal to the number of links in the
 original network. In the new network, two nodes are connected
 if they are formed from links which shared the same node. In
 particular, a node of degree $k$ is converted to a fully
 connected subgraph of $k$ nodes. The definition of distance ensures in particular, 
that the small world effect in the original network persists also 
in the transformed network.\\

Recently the line graph constructed from the scale-free network was discussed
in \cite{nach} analytically and numerically. The results for dense networks 
($<k>$=10, 20 and 30) supported the rule that the transformed network is also 
scale-free. The exponent $\gamma '$ defined by the degree distribution $P(k)\propto k^{-\gamma '}$
of the transformed graph was shown to fulfil the relation $\gamma ' = \gamma -1$, where $\gamma$ was
the same exponent for the initial network. The line graph was also found to be useful 
in the problem of identification of communities in networks \cite{evans}.\\

 The aim of this work is to investigate the clustering coefficient $C$
 in the line graphs transformed from random networks. Our motivation is twofold.
 First, we share the point of view expressed by Mark Newman, as
 reported in our first paragraph. Our former simulations
 \cite{my,myprim} can actually be seen as the same transformation limited to local
 subgraphs and to nodes of degree three. This technique
 allowed us to enhance the clustering coefficient in networks. Second
 issue can be briefly
 expressed as follows. In social networks, an active contact between
 two linked actors excludes at least to some extent
 the contact between each of these two actors and each of their other
 neigbours. If we introduce two states of a link, termed
 for example 'open' and 'closed', then it is clear that there is an
 anticorrelation between links which share the same node;
 two such links cannot be open simultaneously. It can be convenient,
 then, to work on the network of links instead of the
 network of nodes. However, mathematical formulation of many problems
 is expressed in terms of networks of nodes. Here is the
 area where the transformation can be useful.\\
 
 In the next section we describe some examples of the transformed networks.
 Section 3 is devoted to our numerical results. Short discussion closes the text.
 
 \begin{figure}[ht]
 \centering
 {\centering \resizebox*{12cm}{9cm}{\rotatebox{-90}{\includegraphics{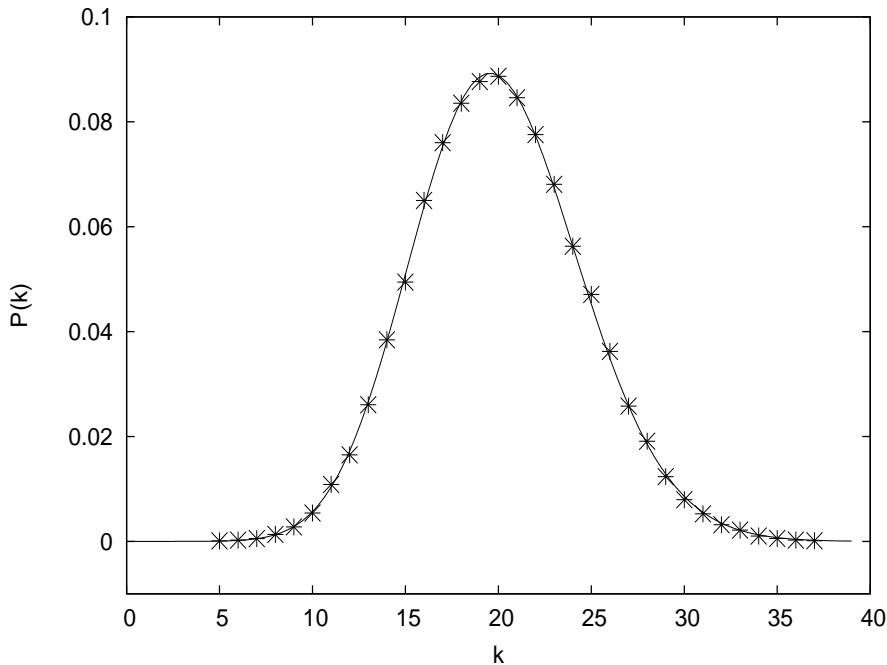}}}}
 \caption{The degree distribution in the network transformed from the
 Erd\"os-R\'enyi network. The stars are numerical results for the initial Erd\"os-R\'enyi network with $<k>=10$ 
and the line comes from the Poisson distribution with mean $\lambda=20$.}
 \label{fig-1}
 \end{figure}

 \begin{figure}[ht]
 \centering
 {\centering \resizebox*{12cm}{9cm}{\rotatebox{-90}{\includegraphics{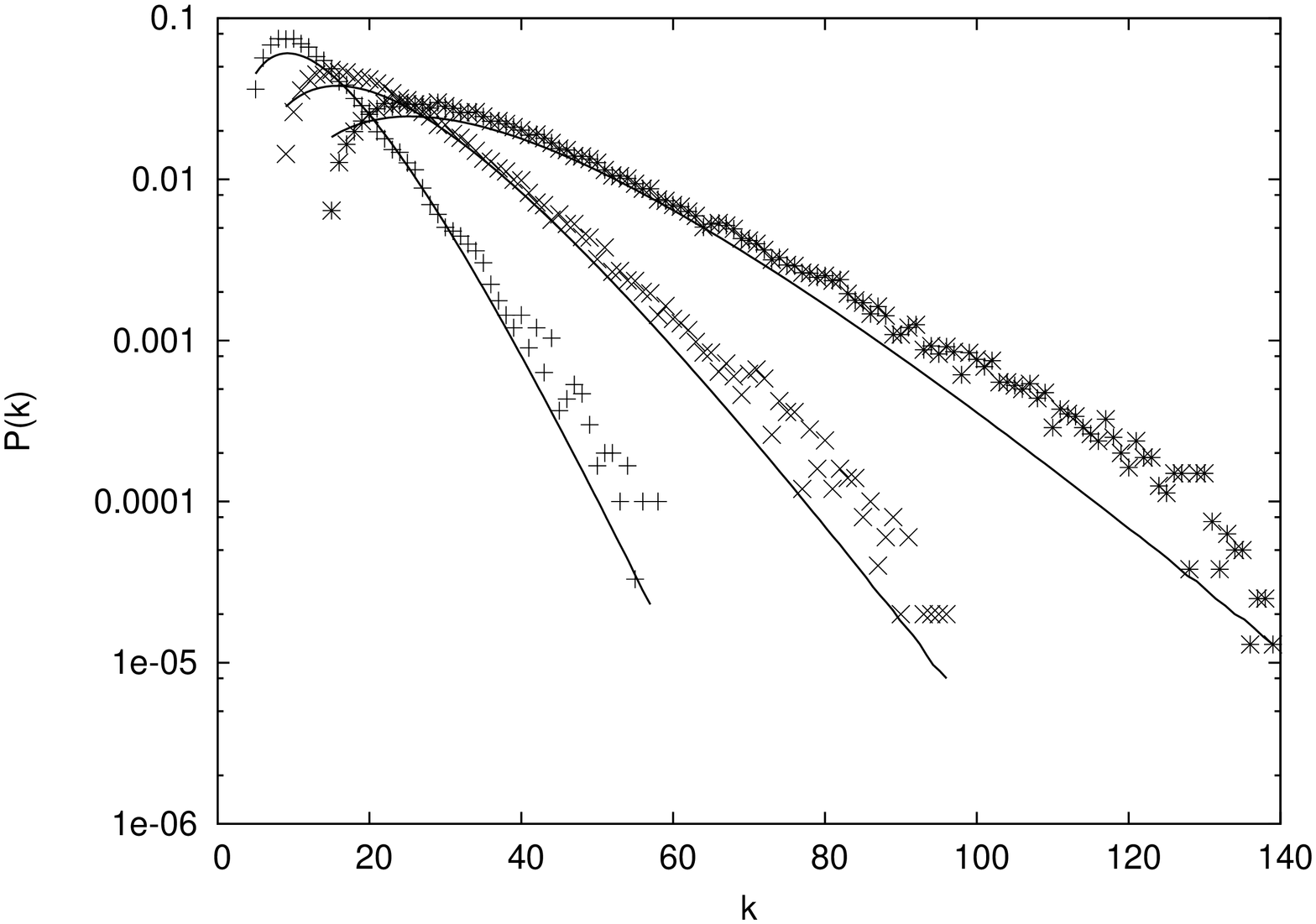}}}}
 \caption{The degree distribution in the networks transformed from the
 exponential networks. The growing parameter $M$=3, 5 and 8  for the curves from left to right.
Numerical results (pluses, X's, stars) are fited with (Eq. 2). The obtained values of $c$ are close to $M/(M+1)$.}
 \label{fig-2}
 \end{figure}

 \section{Examples}

 A chain of $N$ nodes is equivalent to a chain of $N$ links, then under
 action of the transformation $G \to L(G)$ it is converted into itself.
 In a fully connected graph
 of $N$ nodes each pair is connected; there is $N(N-1)/2$ links. After
 the transformation from $G$ to $L(G)$, this is the number of nodes. The transformed
 graph is not fully connected; this can be shown easily for $N=4$.
 There, link $12$ is connected to $13$ and $14$, but not
 to $24$, etc. For arbitrary $N$, each node has $N-1$ neighbours, then
 each link shares its each node with $N-2$ other links.
 In the transformed network, each node has therefore $2(N-2)$
 neighbours. The number of links in the transformed network
 is then $N(N-1)(N-2)/2$. In particular, for $N=4$ in the original graph
 we have 12 links in the transformed graph -
 an octahedron, if links are equally long.\\

 \begin{figure}[ht]
 \centering
 {\centering \resizebox*{12cm}{9cm}{\rotatebox{-90}{\includegraphics{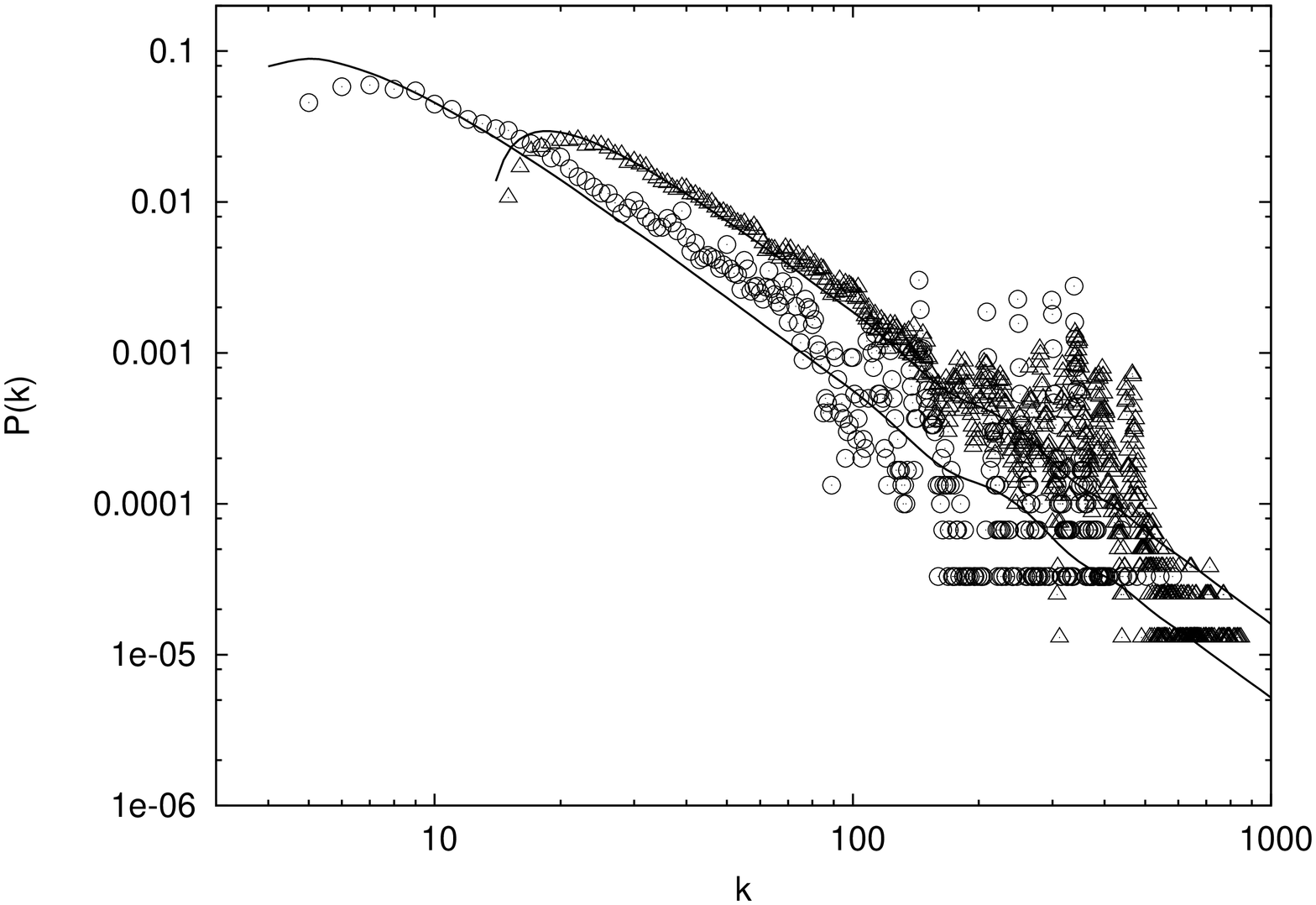}}}}
 \caption{The degree distribution in the networks transformed from the
 scale-free networks for $M=3,8$. The lines come from Eq. 1, with $P(k)\propto k^{-3}$.  }
 \label{fig-3}
 \end{figure}
 
 \begin{figure}[ht]
 \centering
 {\centering \resizebox*{12cm}{9cm}{\rotatebox{-90}{\includegraphics{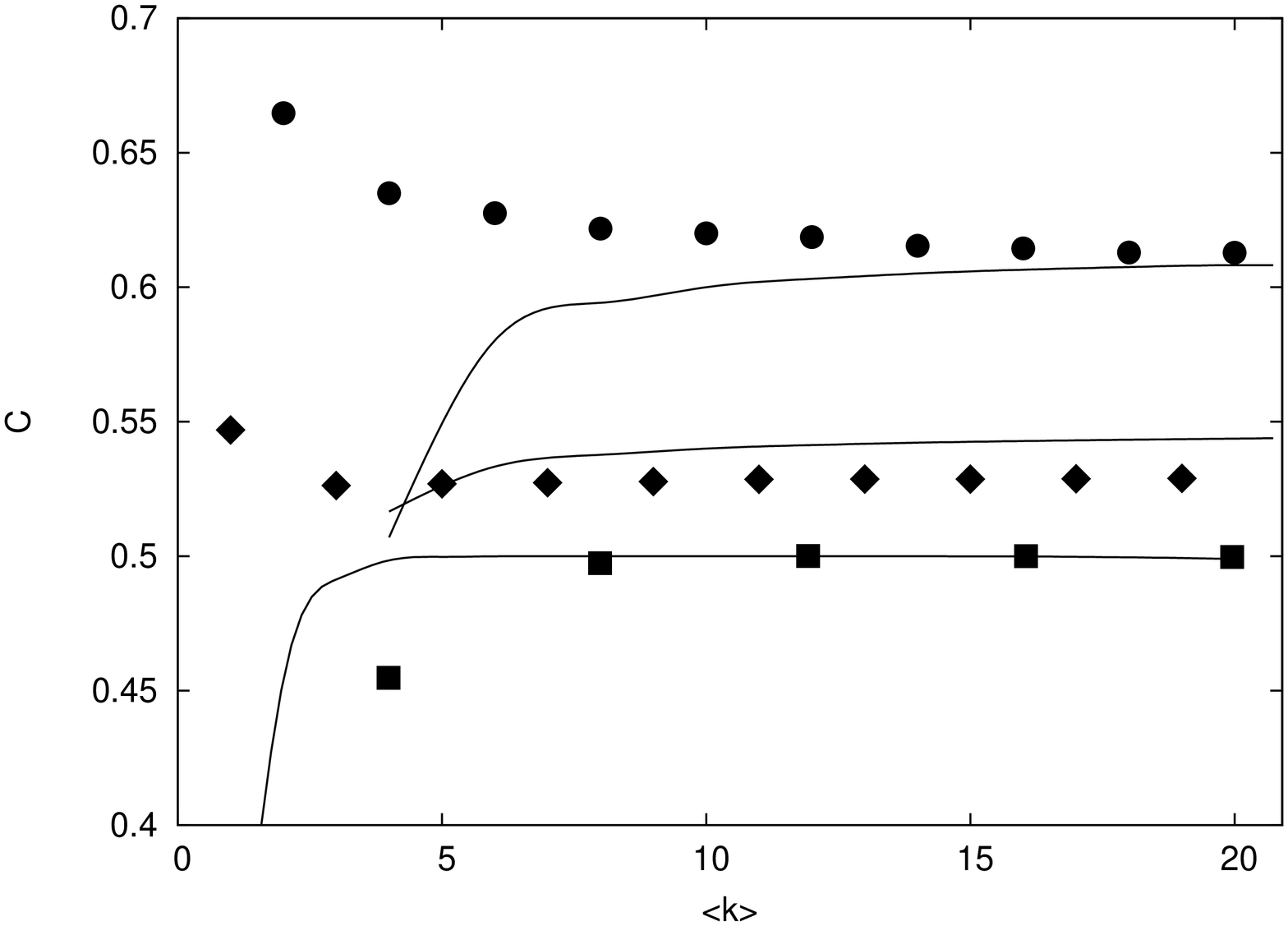}}}}
 \caption{The clusterization coefficient $C$ for the Erd\"os-R\'enyi
 networks against the mean
 degree $<k>$ (squares) and $C$ against the growing parameter $M$ for
 the exponential networks (rhombs) and the scale-free networks (circles). For the 
Erd\"os-R\'enyi network, the continuous line is obtained directly from Eq. 5. For the 
scale-free network we used $P(k)\propto k^{-3}$. For the exponential network, the analytical values of $C$
for large $<k>$ are about 0.545, while the simulation gives $C=0.53$.}
 \label{fig-4}
 \end{figure}
 
 Similar arguments apply to a regular graph, where each node is of the
 same degree, say $k$. Each link joins two nodes
 with $k-1$ other neigbours. Under the transformation $G \to L(G)$, each node is converted into a
 $k$-clique. Obviously, each link in the
 original network joins two nodes; then it contributes - as a node in
 the transformed network - to two cliques.
 Its degree in the transformed network is then $2(k-1)$. The number of
 links in the transformed network is then
 $(Nk/2)\times(k-1)$. In the case of $N=4$ and $k=3$ again a
 tetrahedron is transformed into an octahedron.\\

 Let us consider a network with the degree distribution $P(k)$, where
 is no correlation between degrees of neighboring
 nodes. The above arguments are now as follows. A link in the original
 network joins two nodes with degrees $k_1-1$ and
 $k_2-1$; the considered link is not counted. In the transformed network, this link is a node of degree
 $k_1+k_2-2$. The degree distribution $P_t(k)$ in the
 transformed network is then \cite{dorbi}
 
 \begin{equation}
 P_t(k)=\sum_{k_1,k_2}k_1k_2P(k_1)P(k_2)\delta_{k,k_1+k_2-2}=\sum_{k_1=1}^{k+1}k_1(k-k_1+2)P(k_1)P(k-k_1+2)
 \end{equation}
 
 In the case when $P(k)$ is the Poisson distribution with $<k>=\lambda$ the degree
 distribution $P_t(k)$ for the transformed network is a new Poisson distribution with $<k>=2\lambda$. For
 the geometrical distribution $P(k)=(1-c)c^k$ we get
 
 \begin{equation}
 P_t(k)=\frac{(1-c)^4}{6}(k+1)(k+2)(k+3)c^k
 \end{equation}
 This distribution, when presented as $logP(k)$ against $k$, gives only
 logarithmic deviation from the degree distribution
 $P(k)$ of the original network. For the power function $P(k)\propto
 k^{-\gamma}$ we have no simple result. Some analytical considerations in terms of Polygamma functions can be found in \cite{nach}.\\
 
 Now we consider the clustering coefficient $C$. In a
 fully connected graph of $N$ nodes each link $12$ is
 converted under $G \to L(G)$ to a node of degree $2(N-2)$. Maximal number of
 links between these $2N-4$ neighbours is $(N-2)(2N-5)$.
 The actual number of links within each clique of $(N-2)(N-3)/2$. There
 is also $N-2$ links between nodes converted from
 links which met at the $N-2$ nodes different from nodes $1$ and $2$.
 Then in total we have $(N-2)(N-3)+N-2=(N-2)^2$ links,
 what gives
 
 \begin{equation}
 C=\frac{(N-2)^2}{(N-2)(2N-5)}=\frac{N-2}{2N-5},
 \end{equation}
 the same for each node. For regular graphs of degree $k$ each link is
 converted to a node of degree $2(k-1)$, with
 $(k-1)(2k-3)$ possible links between its neighbours. Neglecting
 triangles built on the considered link, we have only
 $2(k-1)(k-2)/2$ links within cliques. Then the clustering coefficient is
 
 \begin{equation}
 C=\frac{k-2}{2k-3}.
 \end{equation}
 For a network with degree distribution $P(k)$ we have to find an average
 
 \begin{equation}
 C=\sum_{k_1,k_2}k_1k_2P(k_1)P(k_2)\frac{(k_1-1)(k_1-2)+(k_2-1)(k_2-2)}{(k_1+k_2-2)(k_1+k_2-3)}.
 \end{equation}
There, the contribution of pairs of nodes where $k_1+k_2<4$ is zero.

 \section{Numerical calculations and results}
 
 The connectivity matrix for the transformed network is constructed here as
 follows. In the conectivity matrix $C(i,j)$ of the
 original network each unit above the main diagonal means a link. We
 substitute these units by their consecutive numbers:
 $r=$ 1, 2 and so on. Let us call the obtained matrix $R(i,j)$. The
 last number $r_m$ is equal to the number of links in the
 original network; it is then equal also to the size of the transformed
 network. In the conectivity matrix $C_t(i,j)$ of
 this network, elements $i$ and $j$ are connected if their numbers $i$
 and $j$ appear in the matrix $R$ in the same
 row or in the same column.\\
 
 The original Erd\"os-R\'enyi network is generated from $N=10^4$ nodes.
 Then, the number of nodes in the transformed
 network is about $pN^2/2$, where $p$ is the density of links in the
 original network. With $p=10^{-3}$, as in Fig. 1,
 we expect $<k>$ close to $Np=10$ in the original network. The number of
 nodes in the transformed network should be about
 $N^2p/2=5\times 10^4$. In the example presented in Fig. 1 we have
 50147 nodes. The result of the analytical calculation
 made above indicates that the mean degree of the transformed network
 should be equal to 20, what confirms the simulation.\\
  
 The original exponential network is grown from a fully connected
 cluster of $M$ nodes. Each next node is attached
 to randomly selected $M$ different nodes. No preference of attachment
 is applied for the exponential networks. The original
 network has $N=10^4$ nodes, and the transformed network has about $NM$
 nodes. The original network is known to have the exponential
 distribution of node degree; $log(P(k))$ plotted against $k$ is a
 straight line. The degree distributions of the transformed
 network, shown in Fig. 2, seem also to be close to the exponential
 function, as in the original network. The curves shown are obtained from Eq. 2. The size of the 
 transformed network is 29994, 49985 and 79964 for $M=$3, 5 and 8, respectively.\\
  
 To generate the scale-free networks numerically we have only to add the
 preferential attachment; nodes are selected with the probability
 proportional to their degree.The original scale-free network is again $10^4$ nodes. In Fig. 3 we show the degree
distributions for the growing parameter $M=3$ and $M=8$. The plots are not far from straight
lines in the log-log scale. On the contrary to the exponential network, the slope of the 
obtained curves does not depend on the mean degree $<k>=2M$. The overall results agree with those of \cite{nach}.\\

In Fig. 4 we show the comparison of the data on the clustering coefficient $C$, as calculated from direct 
numerical simulations (points) and from Eq. 5 (lines). It appears that the respective plots met when
the mean degree $<k>$ is large enough. This accordance indicates that our model assumption
on the lack of correlations works well for dense networks. As it can be seen in Fig. 4, the largest 
departure of the clustering coefficient $C$ calculated numerically from the analytical values are
found for the exponential networks. This suggests that for these networks, the degree-degree correlations are the largest.\\

\section{Conclusions}

The rule that a given degree distribution is transformed under $G \to L(G)$ into the degree distribution
from the same family has some justification in the transformation itself. Namely, a node of degree $k$
is transformed into a set of nodes of at least the same degree $k$. This means in particular that hubs are
transformed into cliques of hubs, and chains are transformed into chains. For the scale-free networks, our
results on the degree distribution $P(k)$ agree with those of \cite{nach}.\\

Accordance of the results calculated numerically with the analytical formulas means in our case that the 
corrections introduced by degree-degree correlations \cite{past} are relatively irrelevant. We demonstrated that the 
transformation $G \to L(G)$ leads to clustered networks, where the clustering coefficient $C$ is not smaller
than 0.5. This limit value can be lower, if $G \to L(G)$ is applied to selected local subnetworks and not to the 
whole system. The density of the transformed nodes can be used to tune $C$, similarly to \cite{e1,e2,e3,e4,e5,my}.\\

An application of the transformation $G \to L(G)$ to the communication networks needs now a specification of interaction between
links, i.e. between nodes of the transformed network. The task is out of frames of this paper. We hope that the
idea of interaction between links can find applications in networks, where a relation between two nodes
excludes at least partially the relations between one of these nodes and its neighbours. We have in mind trade
networks, sexual networks, decision trees and transport networks, where parallel links are activated along 
the principle 'this or this'. Links can also activate each other, according to 'this, then this'. Examples could 
be found in genetic networks, chains of catalytic reactions and social systems, where a process is simultaneously 
an active agent. 
 
 \section*{Acknowledgements} The authors are grateful to Zdzis{\l}aw Burda and to our Anonymous Referee for helpful remarks
and useful references.
One of the authors (A.M.) is grateful to UNESCO for funds in frames of the Fellowship Programme UNESCO/Poland.
The research is partially supported within the FP7 project SOCIONICAL, No. 231288. The calculations were performed in 
the ACK Cyfronet, Cracow, grants No. MNiSW/SGI3700 /AGH /030/ 2007 and MNiSW/SGI3700/AGH /031/ 2007.

 \end{document}